\documentclass[10pt,twoside,twocolumn,english,showpacs,floatfix,pre]{revtex4}
\usepackage[T1]{fontenc}
\usepackage[latin9]{inputenc}
\usepackage{amsmath}
\usepackage{graphicx}
\usepackage{amssymb}

\makeatletter

\usepackage{babel}

\makeatletter

\begin{document}


\title{Elastic turbulence in a curvilinear channel flow}

\author{Yonggun Jun and Victor Steinberg}

\affiliation{Department of Physics of Complex Systems, Weizmann
Institute of Science, Rehovot, 76100 Israel}

\date{\today}

\begin{abstract}
We report detailed quantitative studies of elastic turbulence in a curvilinear channel flow in a dilute polymer solution of high molecular weight polyacrylamide in a high viscosity water-sugar solvent. Detailed studies of the average and rms velocity and velocity gradients profiles reveal an emergence of the boundary layer associated with the nonuniform distribution of the elastic stresses across the channel. The characteristic boundary width is independent of the Weissenberg number $Wi$ and proportional to the channel width that follows from our early investigations of the boundary layer in elastic turbulence of different flow geometries. The appearance of the characteristic spatial scales of the order of the boundary layer width of both velocity and velocity gradient in the correlation functions of the velocity and velocity gradient fields in a bulk flow suggests that rare and strong parcels of excessive elastic stresses, concentrated in the boundary layer, are ejected into the bulk flow similar to jets observed in passive scalar mixing. And finally, the experimental results show that one of the main predictions of the theory of elastic turbulence, namely the saturation of the normalized rms velocity gradient in the bulk flow of elastic turbulence contradicts to the experimental observations both qualitatively and quantitatively in spite of the fact that the theory explains well the observed sharp decay of the velocity power spectrum. The experimental findings call for further development of theory of elastic turbulence in a bounded container, similar to what was done for a passive scalar problem.
\end{abstract}


\maketitle

\section{Introduction}

The addition of small amount of long polymer molecules into a fluid
makes it elastic and capable to storing elastic stresses that may
strongly alter flow properties \cite{bird}. First of all, the elastic stresses generated by the polymer stretching in the flow becomes the main source of nonlinearity in the polymer solution flow at low Reynolds numbers $Re$. As the result, an elastic instability shows up, when the elastic energy overcomes the dissipation due to polymer relaxation. The ratio of
the nonlinear elastic term to the linear relaxation is defined by
the Weissenberg number $Wi$ \cite{bird}, which is the main control parameter in the problem, and the elastic instability occurs in a shear flow with curvilinear trajectories at $Wi_c\geq 1$. Above the purely elastic instability, a path to a chaotic flow in a
form of irregular flow patterns at $Wi> Wi_c$ was studied in three
flow geometries: Couette flow between cylinders, swirling flow
between two disks, and flow in a curvilinear channel
\cite{Nat,Nat1}. Further increase of $Wi$ at vanishingly small $Re$ leads to the most remarkable
phenomenon discovered recently experimentally \cite{Nat} and then
studied during the last decade in the increasing number of
experimental \cite{Nat1,NJP,teo1,teo2,teo3,jun1} and theoretical \cite{lebedev,berti1,berti2,yatou} papers, namely "elastic
turbulence". It is a spatially smooth and random in time flow, which is driven by strong polymer stretching and resulting elasticity observed at sufficiently large $Wi$ and at vanishingly small $Re$. Many properties of the elastic turbulence
regime, namely statistics of velocity and velocity gradient fields,
spatial and temporal velocity correlation functions and power-law behavior of
velocity power spectrum, a new length scale-the boundary layer width
and its properties, scaling and statistical properties of torque
and pressure fluctuations, in a von Karman swirling flow between two
disks were observed and investigated experimentally and only a few of
them theoretically and numerically \cite{teo3}.

On the other hand, a detailed description of similar properties in a
channel flow is lacking, though elastic turbulence in a curvilinear
channel was used to demonstrate its effectiveness in mixing in
macro- as well as in micro-channels \cite{Nat1,NJP,teo1,teo4,jun2}.
In a 3 mm wide curvilinear channel the longitudinal and transverse
flow velocity components were measured by a laser Doppler anemometer
at the bend $N=12$ near the middle of the half-ring in the middle of
the channel at a fixed value of $Wi$, about twice above the onset to
the elastic instability. The power spectra of the longitudinal and
transversal velocity components show a broad region of an algebraic
decay $f^{-3.3}$ in the frequency domain. The rms velocity of
fluctuations was 0.09$\langle{V}\rangle$ and 0.04$\langle{V}\rangle$ for the longitudinal and transverse components, respectively,
where $\langle{V}\rangle$ is the mean longitudinal velocity component. Since the power spectra were
measured at a point with more than 10 times higher mean than the
characteristic fluctuation velocity, the Taylor hypothesis can be
used to transfer the power-law-decay in the frequency domain to the
wave number $k$ domain as $k^{-3.3}$, which is very close to the
scaling exponent found in the von Karman swirling flow between two
disks in both frequency and wave number domains \cite{Nat,teo3}. Another measurements of the flow velocity were
carried out in the micro-channel of a similar design as the 3 mm
channel but scaled down 30 times compared with the macro-channel
version \cite{teo1,teo4}. In order to define the onset of the
elastic instability, the both longitudinal and transverse velocity
field components were measured by microscopic particle image
velocimetry, micro-PIV, as a function of the pressure drop $\Delta
p$ along the channel. The resulting flow resistance, defined as
$\Delta p/\bar{V}$, as a function of the pressure drop shows sharp
but continuous change in the dependence that determines the instability onset.
Similar effect is observed on the plot of $V_{\theta}^{rms}$ as well as $V_r^{rms}$ versus $\Delta p$
\cite{teo1,teo4}. Temporal dependence of the longitudinal velocity
as well as its correlation function were also measured together with
the velocity correlation time. By using 0.2 $\mu$m fluorescent
particles streamwise vortices were visualized by means of horizontal
confocal scanning microscopy in the middle plane of the channel
\cite{teo1,teo4}. In the recent detailed studies of mixing in a
curvilinear channel of 1 mm$^2$ cross-section using elastic
turbulence, some partial characterization of the velocity field
necessary for quantitative characterization of mixing, namely both
average radial $\langle V_{r}\rangle$ and longitudinal $\langle
V_{\theta}\rangle$ velocity profiles across the channel, dependence
of $\langle V_{\theta}\rangle$ on $Wi$, profiles of rms of
longitudinal $\partial V_{\theta}/\partial r$ and radial $\partial
V_{r}/\partial r$ velocity gradients, was carried out
\cite{jun2}.

The first and currently the only theory of elastic turbulence was developed right away after the first publication of its discovery. The main concern of the theory was to explain the key experimental observation in elastic turbulence, namely the sharp algebraic decay of the velocity power spectrum with the scaling exponent $\delta$ between -3.3 and -3.6 \cite{Nat,Nat1}. Due to the sharp velocity spectrum decay, the
velocity and velocity gradient are both determined mostly by the
integral scale, i.e., the vessel size. It means that elastic
turbulence is essentially a spatially smooth and random in time flow,
dominated by strong nonlinear interaction of a few large-scale
modes. It is the same random flow that occurs
in hydrodynamic turbulence below the dissipation scale and called
Batchelor flow regime \cite{batch}. It is resulted from stretching and folding of elastic stress field, similar to a passive scalar stretching and folding in the Batchelor regime of mixing. A crucial difference between elastic turbulence and passive scalar mixing is that in the case of elastic turbulence the corresponding elastic stress field is not passive but reacts back on the driving velocity field and in such way stabilizes the flow \cite{lebedev,lebedev2}.
There are two aspects of theory of polymer stretching in a flow.
First is a description of statistics of polymer stretching and a coil-stretch
transition in a spatially smooth and random in time flow, and second
is a characterization of the properties of elastic turbulence resulting from the polymer
stretching.

The first aspect requires a microscopic approach to the problem, which provides a quantitative prediction on the coil-stretch transition of a polymer and on a saturation of the polymer stretching in a spatially isotropic random unbounded flow and a detailed prescription to experimentally verify it \cite{lebedev1,chertkov}. This prediction was tested experimentally, and good agreement was found \cite{gerashch,liu1}. The coil-stretch transition has remarkable macroscopic consequence on  a flow: properties of the polymer solution become essentially non-Newtonian and the stretched polymers significantly alter the flow due to their back reaction.
The second aspect, on the other hand, requires a macroscopic description of elastic turbulence, which has been developed by Lebedev et al \cite{lebedev2,lebedev} and is based on polymers with linear elasticity and the feedback reaction on the flow. The theory of
elastic turbulence uses the set of equations for the elastic stress
tensor and velocity fields. Hydrodynamic description of a polymer
solution flow and of dynamics of elastic stresses for linear
polymers is analogous to that of a small-scale fast dynamo in
magneto-hydrodynamics (MHD) and also of turbulent advection of a
passive scalar in the Batchelor regime \cite{lebedev,batch}, though
some significant differences exist. The stretching of the magnetic
lines is similar to the polymer stretching, and the difference with
MHD lies in the relaxation term that replaces the diffusion term in
MHD description, whereas in the passive scalar advection problem the
dynamo effect, i.e. feedback reaction on the flow, is absent. In all three cases the basic physics is the
same, rather general and directly related to the classical Batchelor
regime of mixing: stretching and folding of the passive scalar, magnetic, or stress fields by a
random advecting flow in all three cases.

Theory of elastic turbulence in an unbounded flow of a polymer
solution is based on the following assumptions \cite{lebedev2,lebedev}. (i) A statistically
stationary state occurs due to the feedback reaction of stretched
polymers (or the elastic stress) on the velocity field that leads to
a saturation of the elastic stress $\sigma_p$ and rms of the velocity gradients $(\partial V_i/\partial x_j)_{rms}$
(and so $Wi_{loc}=(\partial V_i/\partial x_j)_{rms}\lambda$, where $\lambda$ is the longest polymer relaxation time) even
for polymers with linear elasticity \cite{lebedev2,lebedev}. The
saturation value in a bulk of elastic turbulence is $Wi_{loc}\simeq
1$ and constant at all $Wi$ above the coil-stretch transition. It is
the key theoretical prediction \cite{lebedev2,lebedev}. (ii) Both
dissipative terms due to viscosity and polymer relaxation, which
appear in the equation for the dissipation of elastic energy
\cite{lebedev2}, are of the same order, i.e.
$\sigma_p/\lambda\sim\eta(\nabla V)^2$ or otherwise
$\sigma_p\lambda/\eta\sim Wi_{loc}^2$, where $\eta$ is the viscosity. Then both assumptions lead to
the following result: the normalized elastic stress
$\sigma_p\lambda/\eta\simeq 1$ and also saturates. These two important
theoretical predictions deserve a stringent experimental test. Indeed, the value of elastic stresses was measured in the recent experiment in microscopic arrangement of a swirling flow \cite{liu} and strong discrepancy was identified. The first prediction on the saturation of $Wi_{loc}$ in a bulk flow of elastic turbulence was examined in a swirling macroscopic flow \cite{teo3} and quantitative disagreement was found. On the other hand, the further theoretical analysis leads to a power-like decaying
spectrum for the elastic stresses and for the velocity field
fluctuations with the exponent $|\delta|>3$ in a good
accord with the experimental results \cite{Nat,Nat1}. The close value of
the exponent in the velocity power spectra decay was also obtained
in the numerical simulations of elastic turbulence based on the
Kolmogorov shear flow of a dilute polymer solution described by the
Oldroyd-B model \cite{berti1,berti2}.

In this paper we provide a complete characterization of the channel
flow in the elastic turbulence regime, present statistics,
correlations and scaling of velocity and velocity gradient fields in the bulk as well as in
the vicinity of the wall, and also properties of the velocity
gradient boundary layer that provide us the possibility also to further test the theoretical prediction about the saturation of $Wi_{loc}$.

The paper is organized as follows. In Sec.
II, the experimental setup is described in details. In Sec. III the
experimental results are presented. Here we first describe in Sec.
IIIA flow structure and experimental determination of the elastic
instability threshold. Then in Sec. IIIB various velocity and velocity gradient profiles
and determination of the boundary layer widths of velocity and velocity gradient and its functional dependence are described. Temporal and spatial correlation functions of velocity and velocity gradients
 and correlation times and lengths as a function of $Wi$ are presented in Sec. IIIC. And finally, statistics of velocity and velocity gradients and structure function scalings are discussed in Sec. IIID. Discussion of the experimental results is given in Sec. IV and conclusions are presented in Sec. V.

\section{Experimental set-up and procedure\label{sec:Experimental-set-up}}

The experiments were conducted in a curvilinear channel of about 1
mm$^2$ square cross-section machined out of plexiglass (lucite). The
width size was chosen to reduce by an order of magnitude the amount
of a working fluid (polymer solution) used in the first experiment
with a channel width of 3 mm \cite{Nat1,NJP} and to increase spatial
resolution of a velocity field in peripheral regions compared with
the second experiment, where a micro-channel of 100 $\mu m^2$ square
cross-section was used \cite{teo1,teo4}. The channel used in the current
experiment contained 63 pairs of identical smoothly connected
half-rings (bends or units) with inner and outer radii of $R_i=1$ mm
and $R_0=2$ mm, respectively, and sufficiently high gap ratio
$d/R_i=1$, the same as in the previous experiments \cite{jun2}, which
was intended to facilitate the onset of the elastic instability at
sufficiently low $Wi$. Here $d=R_0-R_i$ is the channel width. The precise dimensions of the channel were
0.95 mm width at the midplane of the channel, where measurements of
mixing were performed (with $\pm 0.05\mu m$ differences in the width
at the top and bottom of the channel), and 1.025 mm depth. Thus the
entire channel length was approximately 59.4 cm measured along a
channel midplane (see Fig. 1). The channel main body and the lid
sealed with O-ring were squeezed between two stainless steel plates
to seal the channel against leaks and to preserve a flatness along
the channel.

\begin{figure}
\includegraphics[width=8.5cm]{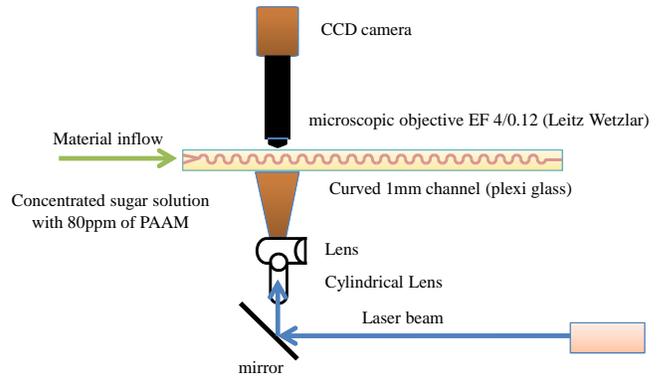}

\caption{The schematic drawing of the experimental setup}

\label{fig1}
\end{figure}

The pulse Nd-YAG laser of the 532 nm wavelength (New Wave Research
Ltd.), which produces pulses of power 30 mJ/5ns with time delay
between the pulses of 10 $\mu$s, was used to conduct particle image
velocimetry (PIV) measurements. Time differences between two consecutive
images was chosen depending on $Wi$ between 0.51 and 1.84 msec, and it
 was synchronized between laser pulses and camera via the control units
 of the PIV laser system.  For visualization
of velocity field measurements, red fluorescent particles (Duke
Scientific Ltd.) of 2 $\mu$m at concentration of 150 ppm were used.
As in the previous setups \cite{Nat1,NJP,jun2}, the channel was
illuminated from a side by laser beam transformed by an appropriate
optical setup containing two cylindrical lenses to a broad light
sheet of a thickness about $\sim 50$ $\mu$m in the observation
region (see Fig. 1). Thinner light sheet causes a permanent damage
to plexiglass. The laser beam sheet produced a thin cut in the 3D
flow, parallel to the top and bottom of the channel at its midplane.
Fluorescent light emitted by the particles in the direction
perpendicular to the beam plane was detected by a charge coupled
device (CCD) via a microscope objective EF 4.0/0.12 (Leitz
Wetzlar) mounted on the plastic tube (guide). The CCD camera
Grasshopper, model Gras-1455M-C (Grey Point Research) of 16 bits
grey scale resolution and with spatial resolution of $1280\times
768$ pixels at up to 30 frames/sec rate was used. Together with the
microscope objective, it provided a PIV spatial resolution of 0.576
$\mu$m/pixel. The size of the PIV image was $1280\times 384$ pixels.
The window size to get velocity vectors was taken $32\times 8$ pixels
that corresponded to $18.4\times 4.6$ ${\mu m}^2$. Total 130 velocity vectors
 in the transverse direction to the flow were obtained from PIV. Thus the velocity gradients were calculated on
$dr=4.6\mu$m. The images were taken at
the rate 10 fps. Since each two consecutive images were transformed
into one velocity field, the final rate was 5Hz of the velocity
field sampling.  Total 1000 velocity field were used for averaging.

The velocity profile in the channel flow of the Newtonian solvent
measured by PIV has been used for calibration. Figure 2 presents the
results of longitudinal velocity profiles $V_{\theta}(r/d)$ across
the channel in a laminar flow of the solvent for various pressure
drops. (We use in the channel middle plane  as the coordinates $z$
and $r$, as longitudinal and transversal to the channel walls,
respectively, and $V_{\theta}$ and $V_{r}$ as the longitudinal and
radial components of the velocity field, respectively.) We visualize
just a part of the channel cross-section to increase a spatial
resolution, since we later on use only the velocity maximum values
and the velocities in a peripheral region close to the wall. Due to curvature in serpentine geometry of the channel, the profiles of the longitudinal velocity are not symmetric relatively to the middle plane in contrast to a straight channel. Figure
3 shows a linear dependence of the maximum values of the velocity
profiles $V_{\theta}^{max}$ (or the discharge) in a laminar flow of
the solvent as a function the pressure drop $\Delta P$ along the
channel, which is used for calibration.
\begin{figure}
\includegraphics[width=8cm]{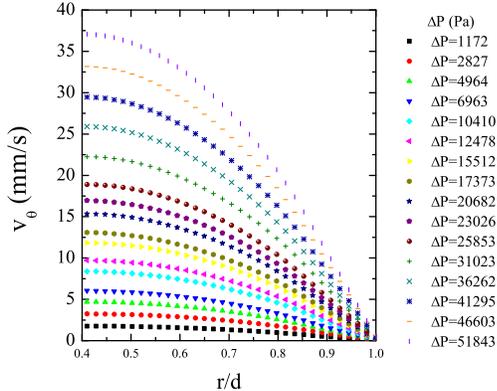}

\caption{Longitudinal velocity profiles across a channel of a
solvent laminar flow in a curvilinear channel at various pressure
drops (starting from small values at bottom to top; $r=0$
corresponds to the inner channel wall). }

\label{fig2}
\end{figure}
\begin{figure}
\includegraphics[width=8.5cm]{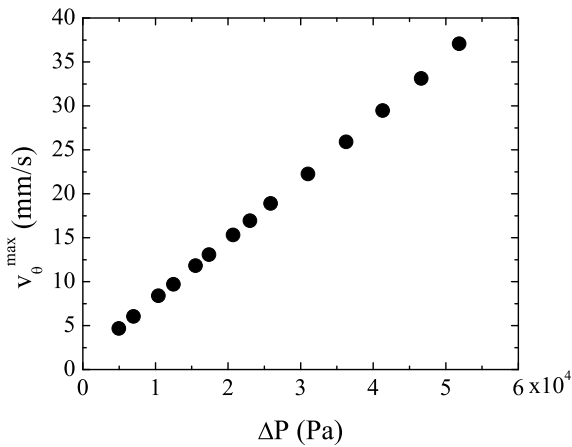}

\caption{Maximum of the velocity profile versus pressure drop along
the channel $\Delta P$ for the solvent of water with 65\% of sucrose
in a laminar channel flow.}

\label{fig3}
\end{figure}

As a working fluid, a 65\% sucrose-water solvent with addition of
1\% NaCl, 250 ppm of $NaN_3$, and 80 ppm by weight high molecular
weight polyacrylamide (PAAm) ($M_w=18$ Mda, supplied by
Polysciences), was used. It was prepared from a master water
solution contained 3000 ppm of PAAm, 1\% NaCl, 250 ppm of $NaN_3$,
and 3\% of iso-propanol \cite{jun2}. The viscosity of the solvent
$\eta_s=113.8$ $mPa\cdot s$ and the polymer solution $\eta=137.6$
$mPa\cdot s$ were measured at $ 22\pm 0.5$ $^\circ$C, the same
temperature kept during the experiment. The longest polymer
relaxation time measured by the stress relaxation method was
$\lambda=11.5$ sec and was found to be independent of the shear rate
\cite{liu} (it differs from values used in Ref. \cite{teo1,teo4},
where the polymer relaxation time was measured by the small
amplitude oscillation method). The solution density was
$\rho=1.303\pm 0.03$ g cm$^{-3}$. The inflow of a polymer solution
into the channel was generated via two inlets by a compressed
nitrogen gas at pressures between 0.35 and 8.5 psi depending on $Wi$. The gas pressure was regulated and
measured via a regulated pressure gauge with a precision of $\pm 5$
Pa. The compressed gas was fed into two plastic cylinders containing
a working polymer solution. The cylinders were connected to the
channel inlets by Tygon tubes with inner diameter of 1.0 mm,
sufficiently large to prevent a possibility of an elastic
instability already in the tubes. The Weissenberg numbers
$Wi=(2V_{\theta}^{max}/d)\lambda$ reached in the experiment by a
such arrangement were in the range from 0 to 950. The maximum
$Re=V_{\theta}^{max}d\rho/\eta$ was $\sim 0.54$, so that the
inertial effects in the flow were negligible, as was already shown
in the earlier experiments \cite{Nat1,NJP,teo1,teo4}.

\section{Properties of elastic turbulence\label{sec:Properties-of-elastic}}

\subsection{Flow structure and experimental determination of the elastic instability threshold
\label{sub:Flow-structure}}

We searched experimentally for the onset of
the elastic instability in our channel with the aspect ratio $d/R_i=1$. In order
to figure out the range of the elastic turbulence region, we plot the values of
$\langle V_{\theta}\rangle$ at $r/d=0.5$ and $V_{\theta}^{rms}$ as
a function of $Wi$ in Figs. 4 and 5. From the both plots one
concludes  that the elastic instability onset corresponds to the change in the slope that occurs at $Wi_c\simeq 200$ on the both plots. Similar change in the slope takes place in the dependence of $V_r^{rms}$ on $Wi$ at about the same value of $Wi_c$ though the data is noisier (Fig. 6). An additional change in functional behavior of $V_{\theta}^{rms}$ and $V_r^{rms}$ versus $Wi$ occurs at $Wi\approx 350-400$, where the transition region to elastic turbulence is ended. In all three plots in Figures 4, 5, and 6 the arrows indicate the onset of the elastic instability and the end of the transitional region towards elastic turbulence. Thus, the range of developed elastic turbulence can be considered
for $Wi$ from about 350 up to 950.

\begin{figure}
\includegraphics[width=8.5cm]{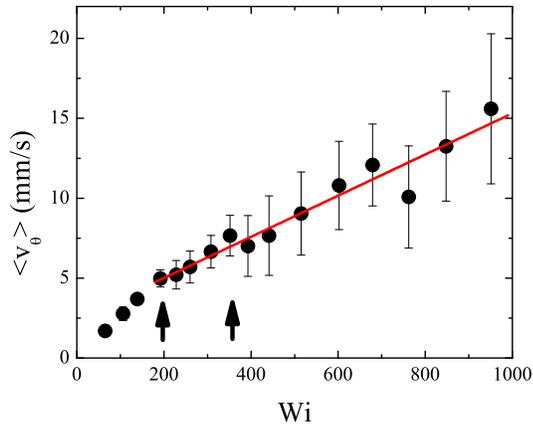}

\caption{The mean longitudinal velocity at $r/d=0.5$ as a function
of $Wi$. Arrows indicate the elastic instability onset and the ending of  a transitional regime to elastic turbulence, respectively. }

\label{fig4}
\end{figure}
\begin{figure}
\includegraphics[width=8.5cm]{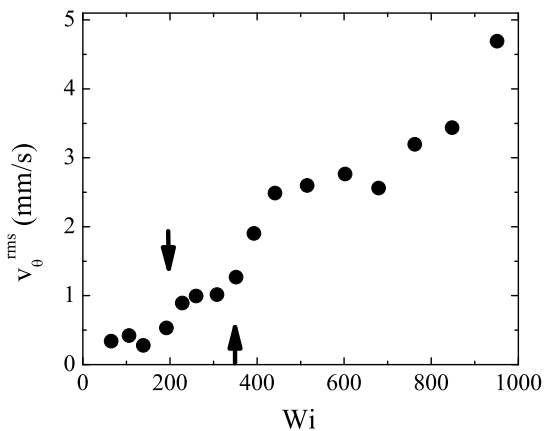}

\caption{ Root mean square of the longitudinal velocity
$V_{\theta}^{rms}$ versus $Wi$. Arrows indicate the elastic instability onset and the ending of  a transitional regime to elastic turbulence, respectively.}

\label{fig5}
\end{figure}
\begin{figure}
\includegraphics[width=8.5cm]{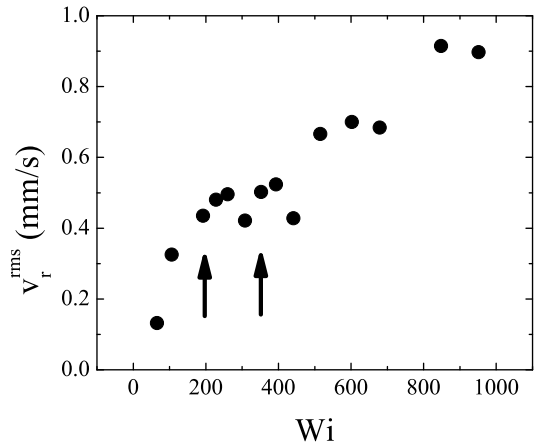}

\caption{ Root mean square of the transversal velocity $V_{r}^{rms}$
versus $Wi$. Arrows indicate the elastic instability onset and the ending of  a transitional regime to elastic turbulence, respectively.}

\label{fig6}
\end{figure}

Using the results of the PIV measurements one calculates the various components of velocity gradient field as a function of r at different values of $Wi$. Figure 7 presents the data for $Wi_{loc}\equiv(\partial
V_{\theta}/\partial r)_{rms}\lambda$ versus $Wi$ averaged over $r/d$ from 0.2
till 0.5 in the bulk flow. In the range of elastic turbulence for $Wi$ from 350 till 950 the data in log-linear presentation in Fig. 7 are fitted by a
linear fit, so $Wi_{loc}=82.2\exp(Wi/919.2)$. Figure 8 show $(\partial V_{r}/\partial r)^{rms}$ averaged over $r/d$ from 0.2 till 0.5 in the bulk flow region as a function of $Wi$. Here the data can be also fitted linearly by $(\partial
V_{r}/\partial r)^{rms}=5.64+0.055Wi$ in the
elastic turbulence range. All PIV data were averaged over 1000
velocity fields to get both average velocity and its rms values as
well as the velocity gradient components.

\begin{figure}
\includegraphics[width=8.5cm]{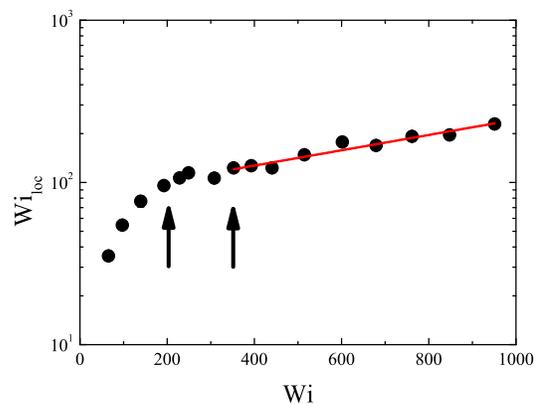}

\caption{The local Weissenberg  number $Wi_{loc}=(\partial
V_{\theta}/\partial r)^{rms}\lambda$ averaged over $r/d$ from 0.2
till 0.5 as a function of $Wi$. Arrows indicate the elastic instability onset and the ending of  a transitional regime to elastic turbulence, respectively.}

\label{fig7}
\end{figure}
\begin{figure}
\includegraphics[width=8.5cm]{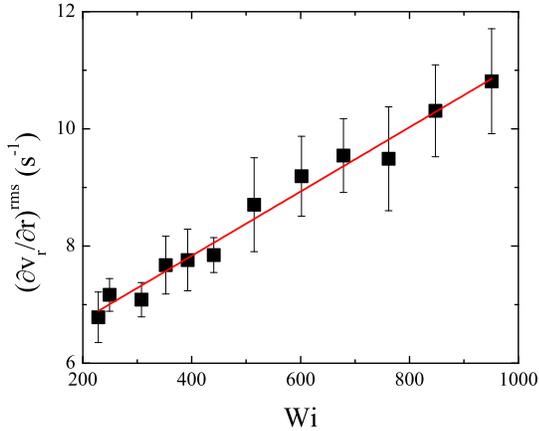}

\caption{$(\partial V_{r}/\partial r)^{rms}$ averaged over $r/d$
from 0.2 till 0.5 versus $Wi$. Solid line is a linear fit to the
data $(\partial V_{r}/\partial r)^{rms}=(5.64\pm 0.15) +(0.055\pm
0.002)Wi$ in the elastic turbulence regime from $Wi=350$ till 950.}

\label{fig8}
\end{figure}

\subsection{Velocity and velocity gradient profiles and the boundary layer
problem\label{sub:Velocity-and-boundary}}

As it is demonstrated above, a channel flow of a polymer solution at sufficiently large $Wi$
above the elastic instability threshold $Wi_c$ is chaotic with large velocity
fluctuations. Figure 9 shows the profiles of the average longitudinal velocity
 $\langle V_{\theta}\rangle (r/d)$ across the channel for
various values of $Wi$ from below the elastic instability and up to
the highest values. A boundary layer characterized by a sharp drop
of $\langle V_{\theta}\rangle (r/d)$ at about $r/d\leq 0.05$ is
clearly seen at $Wi$ above 192. A full profile of $\langle V_{\theta}\rangle (r/d)$
across the channel at $Wi=951$ is presented in Fig. 10, where both boundary layers are clearly identified.
\begin{figure}
\includegraphics[width=8.5cm]{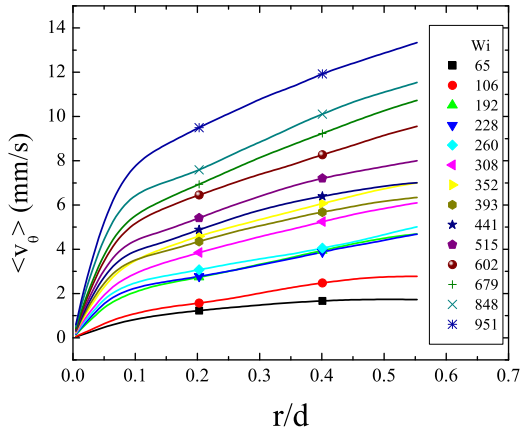}

\caption{Mean longitudinal velocity profiles across the channel of a
polymer solution flow for various $Wi$ (starting from small values
at bottom to top; $r=0$ corresponds to the inner channel wall). }

\label{fig9}
\end{figure}
\begin{figure}
\includegraphics[width=8.5cm]{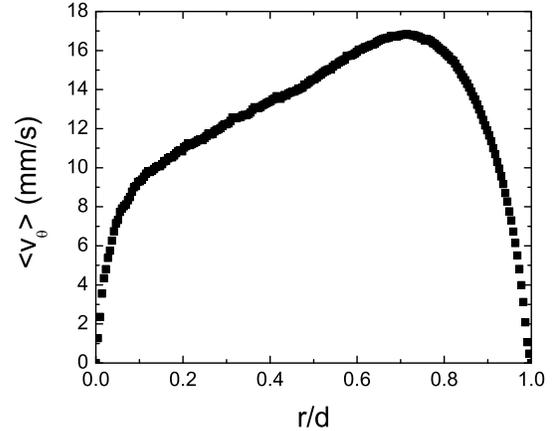}

\caption{The entire mean longitudinal velocity profile across the
channel of a polymer solution flow for $Wi=951$.  }

\label{fig10}
\end{figure}

From the same velocity  measurements the profiles of the average
transversal velocity $\langle V_r\rangle$, and the rms fluctuations
of the longitudinal $V_{\theta}^{rms}$ and transversal $V_{r}^{rms}$
components of the velocity for various $Wi$ are also obtained (see Figs.
11,12,13).

\begin{figure}
\includegraphics[width=8.5cm]{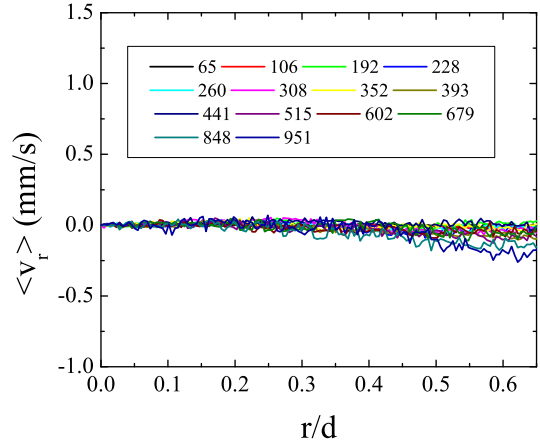}

\caption{Mean transversal velocity profiles across the channel of a
polymer solution flow for various $Wi$. }

\label{fig11}
\end{figure}

\begin{figure}
\includegraphics[width=8.5cm]{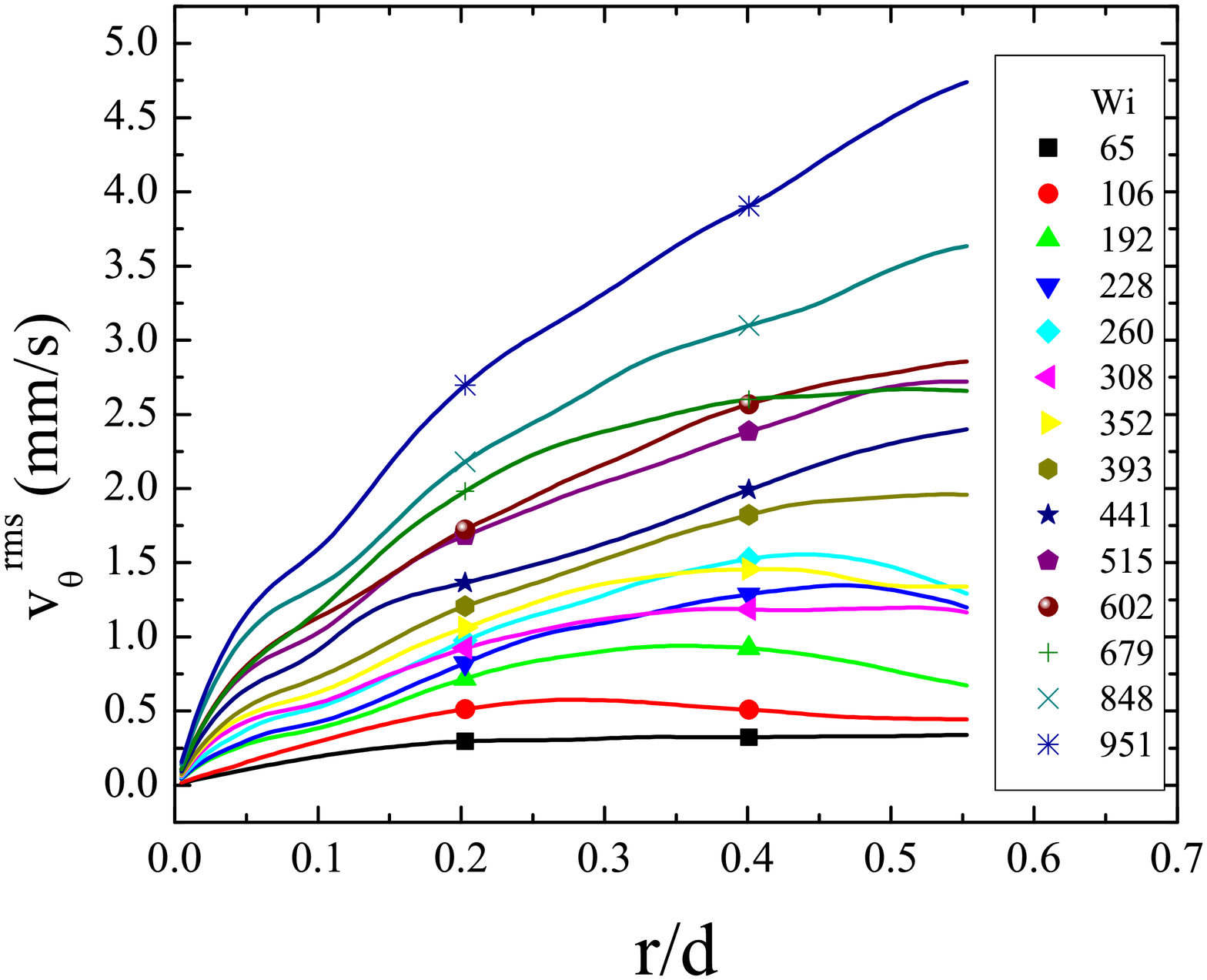}

\caption{RMS fluctuations of the longitudinal velocity profiles across the channel
of a polymer solution flow for various $Wi$
(starting from small values at bottom to top; $r=0$ corresponds to
the inner channel wall). }

\label{fig12}
\end{figure}
\begin{figure}
\includegraphics[width=8.5cm]{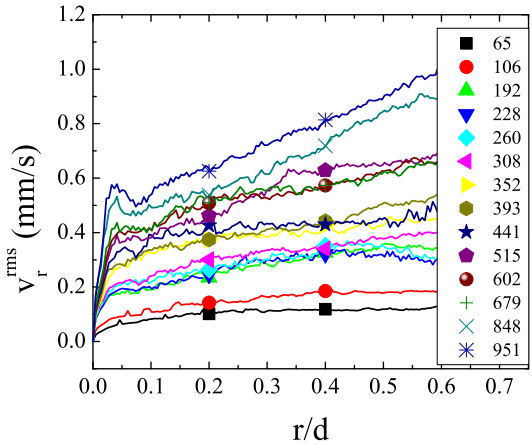}

\caption{Rms fluctuations of transversal velocity profiles across
the channel of a polymer solution flow for various $Wi$ (starting
from small values at bottom to top; $r=0$ corresponds to the inner
channel wall). }

\label{fig13}
\end{figure}

First, by comparing Figs. 9 and 11 one finds that $\langle V_r\rangle$ is close to zero in the limit of error bars across the channel and $\langle V_{\theta}\rangle $ exceeds it by more
than two orders of magnitude. Whereas
rms fluctuations of the longitudinal and transversal components are
of the same order of magnitude that follows from Figs. 12 and 13.
Figure 14 shows $(\partial V_{\theta}/\partial r)^{rms}$ versus $r/d$
at various $Wi$ in the whole range of its variation.  Analogously, the component $(\partial
V_{r}/\partial r)^{rms}$ as a function of $r/d$ is plotted for
various $Wi$ in Fig. 15. As before, all PIV data were averaged over
1000 velocity fields to get both average velocity and its rms values
as well as the velocity gradient components.

\begin{figure}
\includegraphics[width=8.5cm]{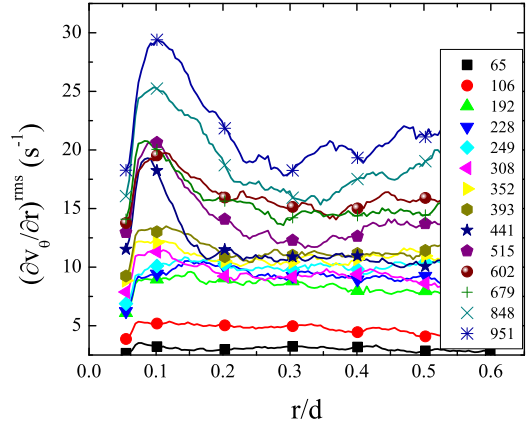}

\caption{$(\partial V_{\theta}/\partial r)^{rms}$ versus $r/d$ for
various $Wi$ from bottom to top. }

\label{fig14}
\end{figure}

\begin{figure}
\includegraphics[width=8.5cm]{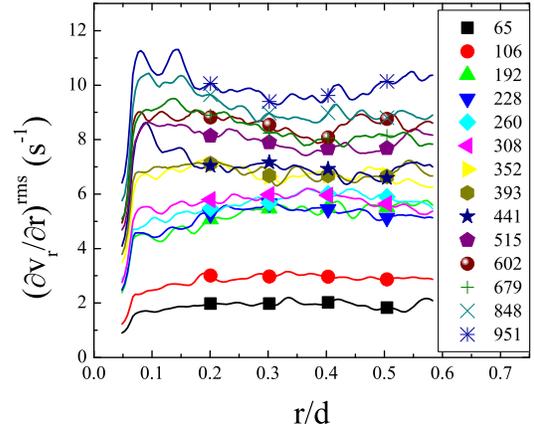}

\caption{$(\partial V_{r}/\partial r)^{rms}$  versus $r/d$ for
various $Wi$ from bottom to top.}

\label{fig15}
\end{figure}

Second, we present in Figs. 16, 17 time dependencies of $V_{\theta}(t)$ and $V_r(t)$ for fixed value of $r/d=0.5$ in the bulk flow at several values of $Wi$ and on time interval much larger than $\lambda$. In the elastic turbulence regime at $Wi>350$ strong fluctuations in the both components of the velocity are observed.

\begin{figure}
\includegraphics[width=8.5cm]{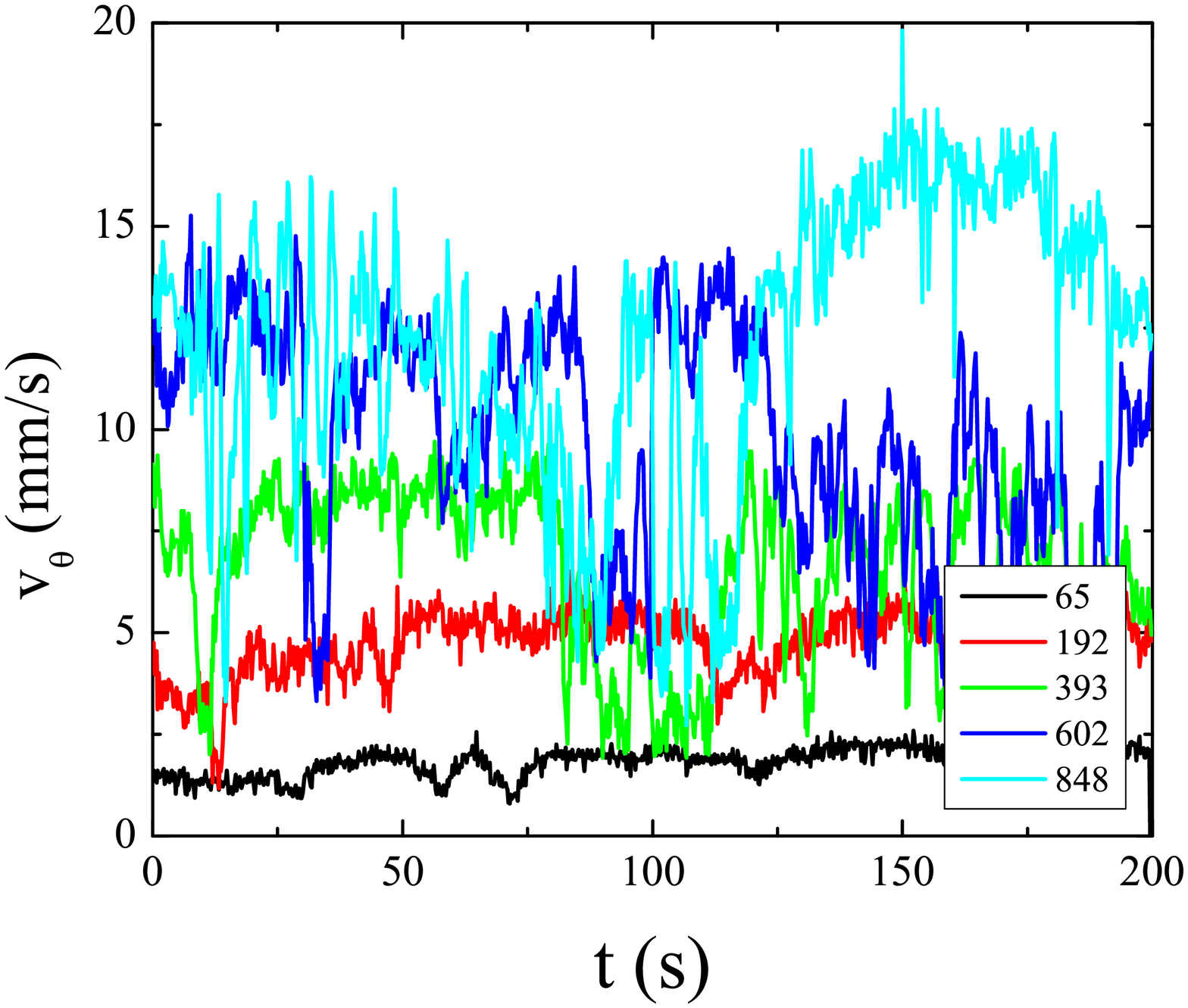}

\caption{Time series of the longitudinal velocity at $r/d=0.5$ for various $Wi$.}

\label{fig16}
\end{figure}

\begin{figure}
\includegraphics[width=8.5cm]{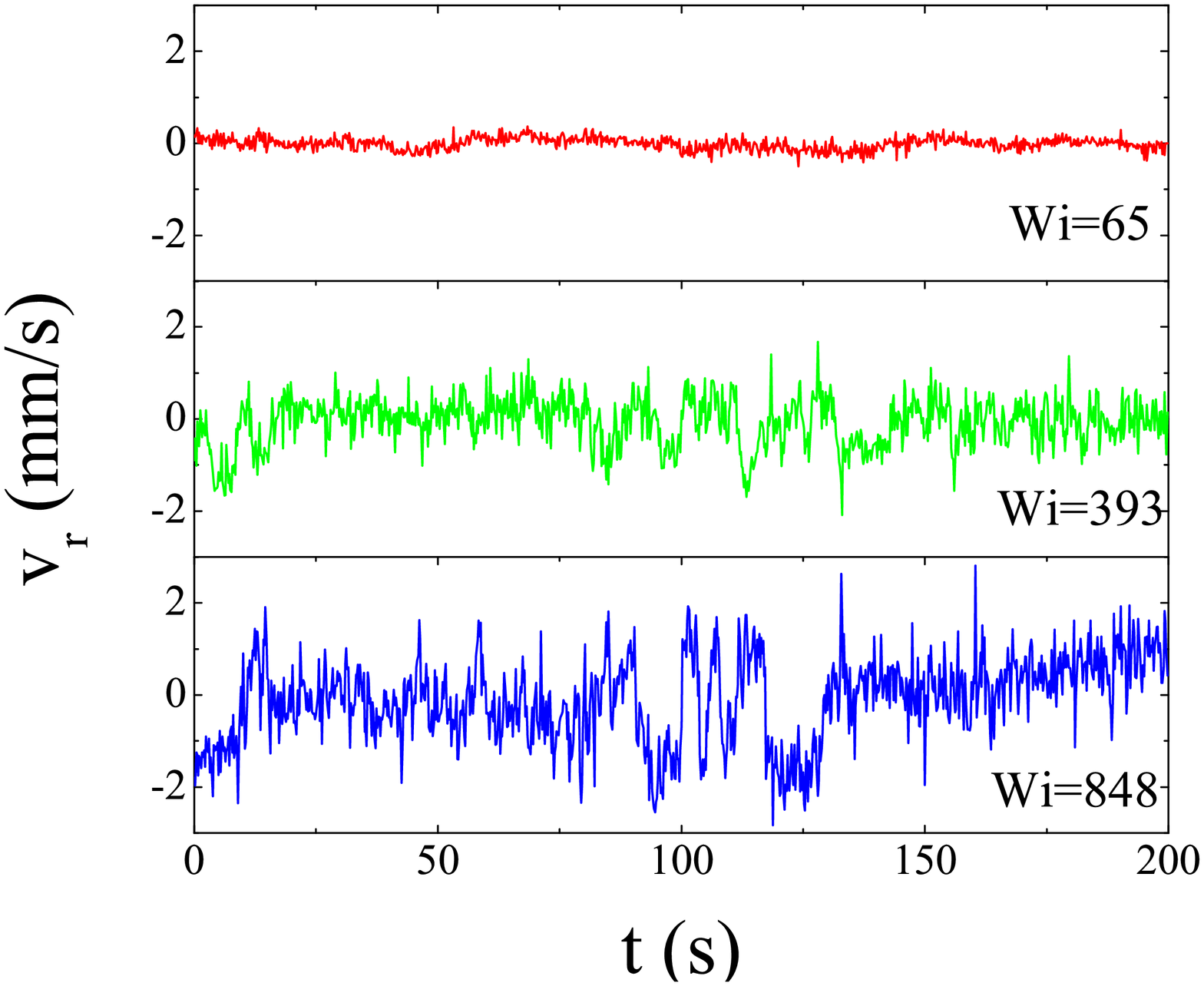}

\caption{Time series of the transversal velocity at $r/d=0.5$ for various $Wi$.}

\label{fig17}
\end{figure}

Clearly identified boundary layer regions near the wall in the both
average velocity and rms of fluctuations of both components of the
velocity and velocity gradient profiles across the channel in Figs. 9,10,14,15 can be better study
after subtraction from each velocity profile a linear part of the
profile found in the bulk region with the slope $d\langle
V_{\theta}\rangle/dr$ (see Figs. 9 and 10) and then scaling its value to
unity. As the result, all longitudinal velocity profiles at $Wi$ in the range of
elastic turbulence above 350 collapse on a  single curve
demonstrating independence of the boundary layer width on $Wi$
(Figs. 18,19). This property of the boundary layer agrees with
that observed in the swirling flow \cite{teo3}. The boundary layer width
is also independent of $Wi_{loc}$
(Fig. 19, inset).
\begin{figure}
\includegraphics[width=8.5cm]{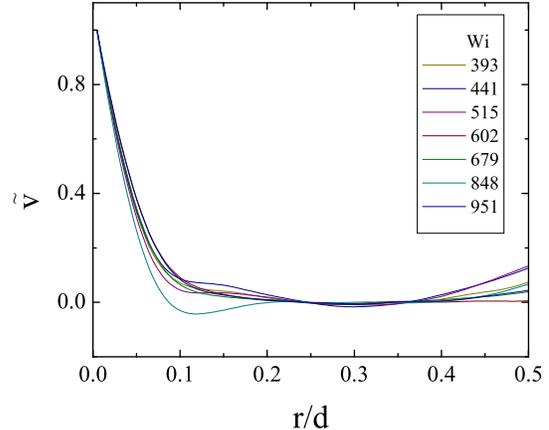}

\caption{Normalized and scaled down $\langle\tilde{V}_{\theta}\rangle (r/d)$
 velocity profiles at various $Wi$.}

\label{fig18}
\end{figure}

\begin{figure}
\includegraphics[width=8.5cm]{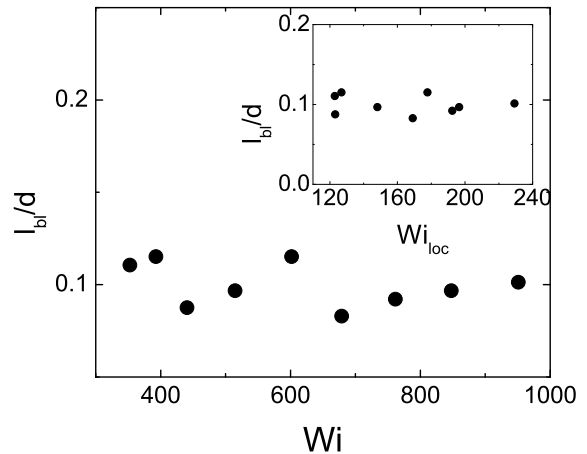}

\caption{Velocity boundary layer width $l_{bl}$ versus $Wi$. Inset:
Velocity boundary layer width $l_{bl}$ versus $Wi_{loc}$.}

\label{fig19}
\end{figure}

 And finally, we present in Fig. 20 the dependencies of $(\partial
V_{\theta}/\partial r)^{rms}$ in the bulk and and its peak values in
the boundary layer regions as a  function of $Wi$ taken from the
plots in Figs. 7 and 14. The peak values of $(\partial V_{\theta}/\partial
r)^{rms}$ in the boundary layer exceed those in the bulk up two
times only compared to at least an order of magnitude in the
swirling flow \cite{teo3,jun3}, probably due to the limiting spatial resolution of
micro-PIV. The latter smears out the peak value of $(\partial
V_{\theta}/\partial r)^{rms}$.

\begin{figure}
\includegraphics[width=8.5cm]{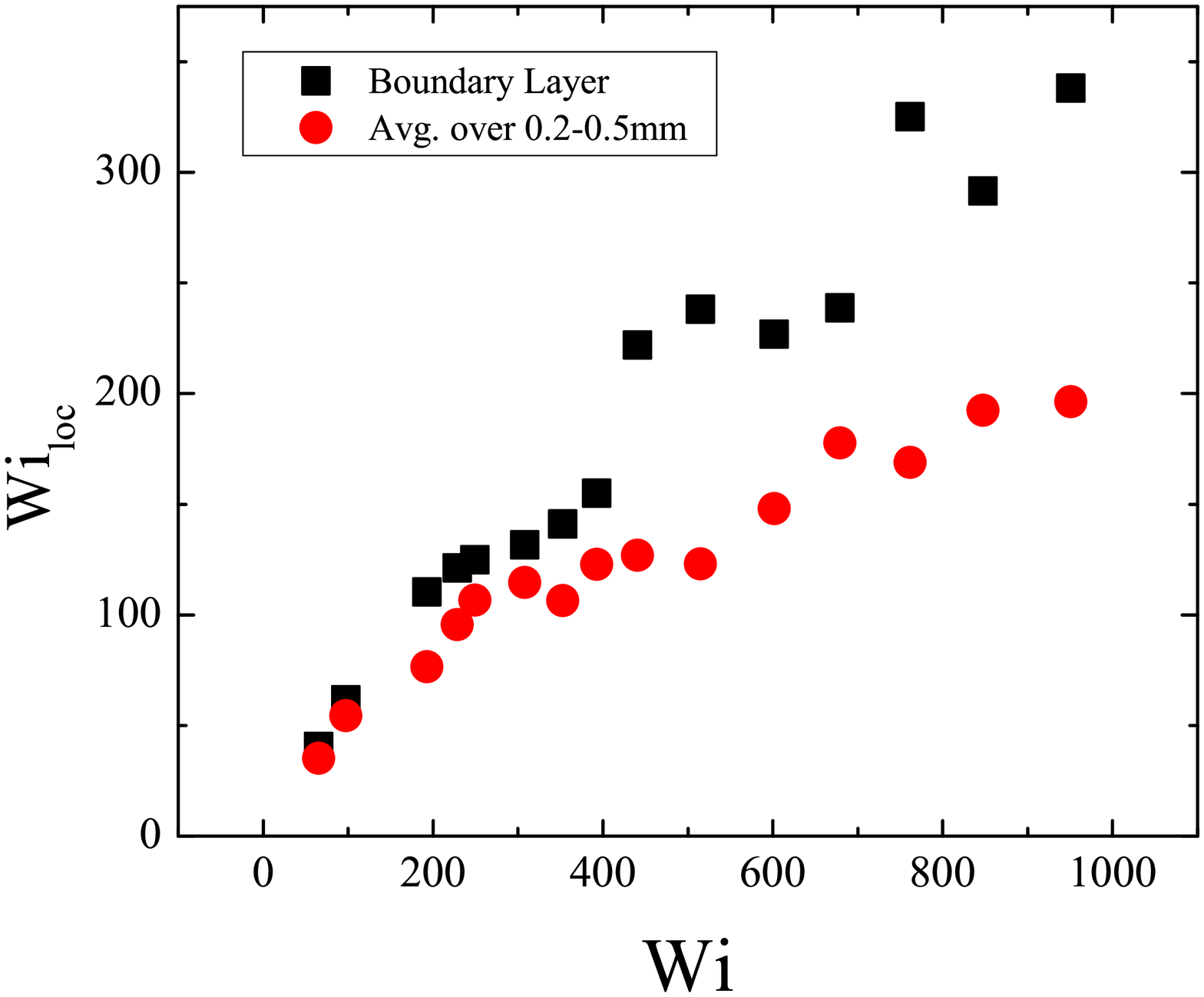}

\caption{ The peak values of $(\partial V_{\theta}/\partial
r)^{rms}$ in the bulk and boundary layer versus $Wi$.}

\label{fig20}
\end{figure}

\subsection{Temporal and spatial correlation functions of both velocity components and their gradients. \label{sub:Power spectra}}

We studied both temporal and spatial (across the channel) correlation
functions of the both velocity components. Figure 21 shows temporal
correlation functions of the longitudinal and transversal velocity components $C(\tau)=\langle \delta V(\tau)\delta
V(0)\rangle/\langle(\delta V(0))^2\rangle$, where $\delta
V_{\theta}(\tau)\equiv V_{\theta}(\tau)-\langle V_{\theta}\rangle$,
and $\delta V_r(\tau)\equiv V_r(\tau)-\langle V_r\rangle$ at several
$Wi$ values in the elastic turbulence regime taken at the bend $N=42$.
The corresponding correlation time calculated as $\tau_{
corr}=\int{tC_{\tau}(t)dt}/\int{C_{\tau}(t)dt}$ for the both
velocity components as a function of $Wi$ are
presented in Fig. 22. As can be seen in Fig. 22, the correlation time for $\delta V_{\theta}(\tau)$ is considerably larger than for for $\delta V_r(\tau)$.

The spatial correlation functions
$C(r/d)= \langle \delta V(r/d)\delta
V(0)\rangle/\langle(\delta V(0))^2\rangle$ and the corresponding
correlation lengths for the both velocity components calculated as
$l_{corr}/d=\int{(r/d)C(r/d)dr/d}/\int{C(r/d)dr/d}$ are shown for several
$Wi$ in Figs. 23 and 24. Here $\delta V_{\theta}(r/d)\equiv
V_{\theta}(r/d)-\langle V_{\theta}\rangle$ and $\delta V_r
(r/d)\equiv V_r(r/d)-\langle V_r\rangle$. The
correlation lengths for both components are
rather close and independent of $Wi$. Using the Taylor hypothesis one can compare the correlation lengths along the flow with those across the flow. As follows from the data in Fig. 22, the correlation time for the longitudinal component of velocity is about $\tau_{corr}/\lambda\simeq 4$ that gives for, let say, $Wi=600$ the correlation length along the flow $\Gamma/d=\tau_{corr}\langle V_{\theta}\rangle\simeq 370$ that is about 2000 times larger than $l_{corr}/d$. Similarly, the correlation functions $\widetilde{C_{\theta}}(r/d)$ and $\widetilde{C_{r}}(r/d)$ as well as the corresponding correlation lengths $\widetilde{l_{\theta,corr}}/d$ and $\widetilde{l_{r,corr}}/d$ of
the both velocity gradients $\partial V_{\theta}/\partial r$ and $\partial V_r/\partial r$ are also calculated, and the results for different $Wi$ were presented in Figs. 25 and 26. First, the correlation functions have minimum at $r/d\approx 0.06$, and second, the correlation lengths are up to an order of magnitude shorter than for the velocities (see Fig. 26 versus Fig. 24) and grow with $Wi$.

\begin{figure}
\includegraphics[width=8.5cm]{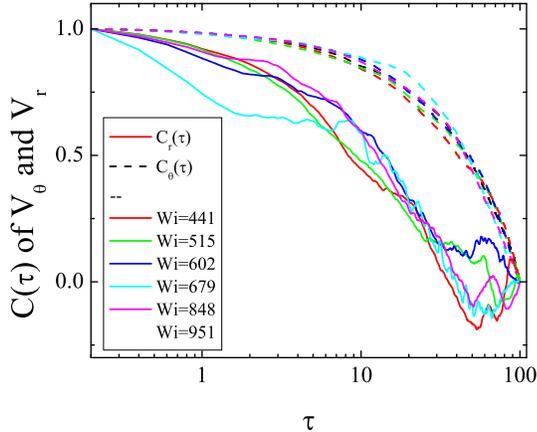}

\caption{Temporal correlation functions $C_{\theta}(\tau)$ of $V_{\theta}(t)$ and temporal correlation functions $C_{r}(\tau)$ of $V_{r}(t)$, respectively, at several $Wi$ above $Wi_c$ taken at the bend $N=42$.}

\label{fig21}
\end{figure}

\begin{figure}
\includegraphics[width=8.5cm]{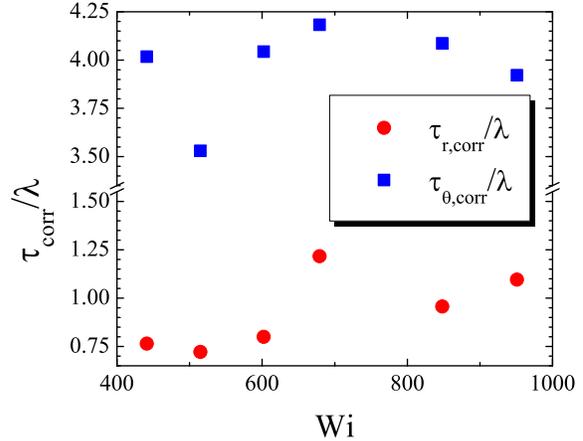}

\caption{Normalized correlation times $\tau_{\theta,corr}/\lambda$ and $\tau_{r,corr}/\lambda$ versus $Wi$ for both $V_{\theta}$ and $V_{r}$, respectively.}

\label{fig22}
\end{figure}

\begin{figure}
\includegraphics[width=8.5cm]{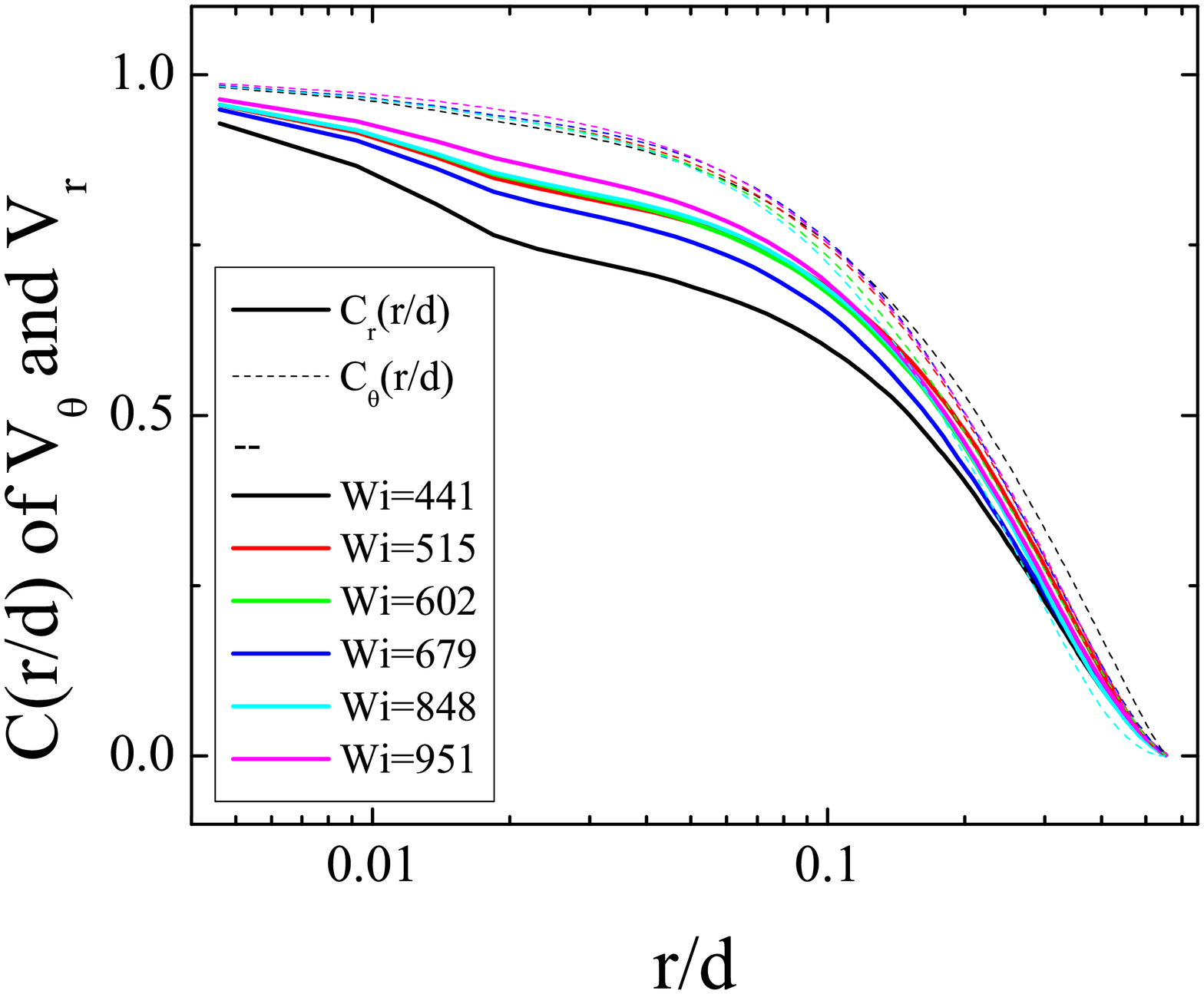}

\caption{Spatial correlation functions $C_{\theta}(r/d)$ of $V_{\theta}(r/d)$ and spatial correlation functions $C_{r}(r/d)$ of $V_{r}(r/d)$, respectively, at several $Wi$ above $Wi_c$ taken at the bend $N=42$.}

\label{fig23}
\end{figure}

\begin{figure}
\includegraphics[width=8.5cm]{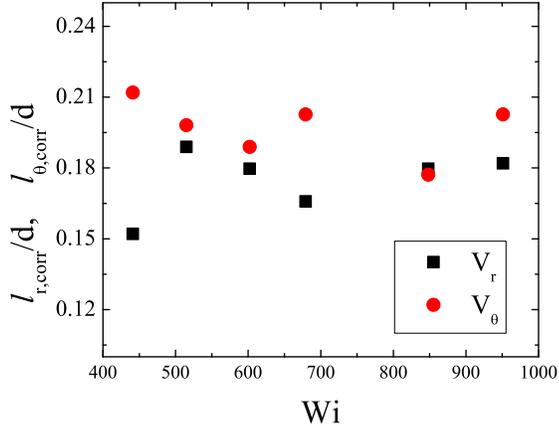}

\caption{Normalized correlation length $l_{\theta,corr}/d$ and $l_{r,corr}/d$ versus $Wi$ for both $V_{\theta}$ and $V_{r}$, respectively.}

\label{fig24}
\end{figure}

\begin{figure}
\includegraphics[width=8.5cm]{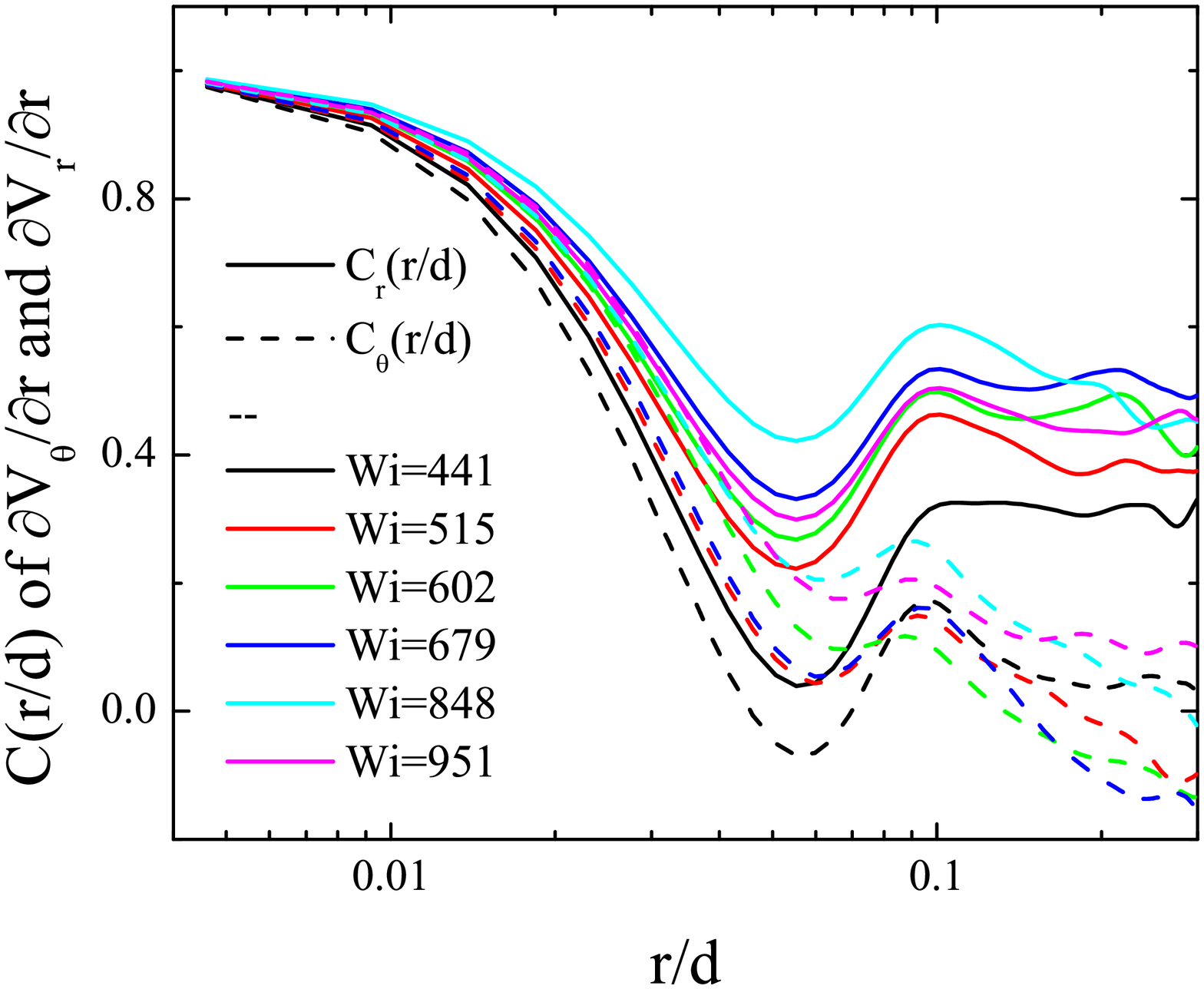}

\caption{Spatial correlation functions $\widetilde{C_{\theta}}(r/d)$ of $\partial V_{\theta}/\partial r$ and spatial correlation functions $\widetilde{C_{r}}(r/d)$ of $\partial V_{r}/\partial r$, respectively, at several $Wi$ above $Wi_c$ taken at the bend $N=42$. }

\label{fig25}
\end{figure}

\begin{figure}
\includegraphics[width=8.5cm]{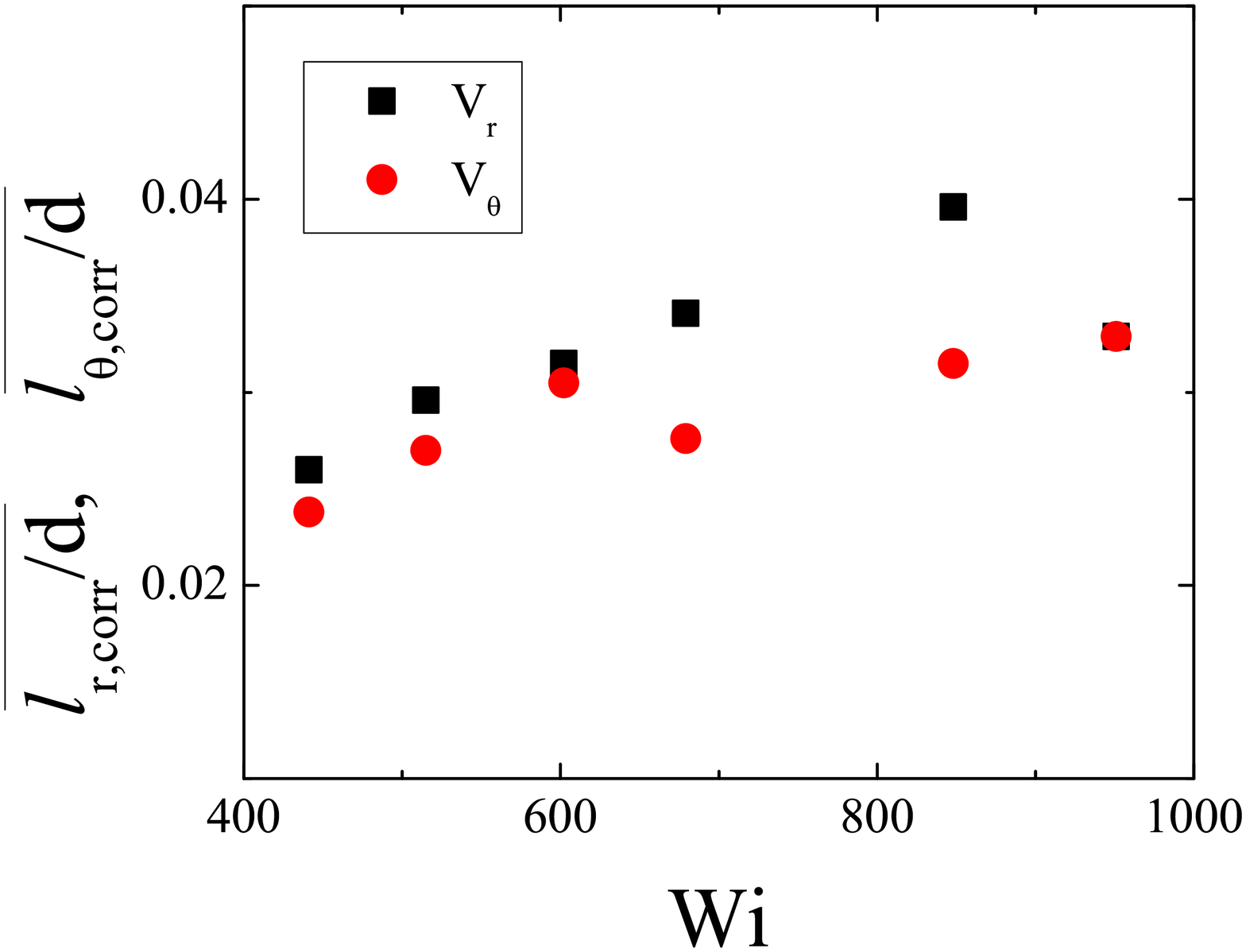}

\caption{Normalized correlation length $\widetilde{l}_{\theta,corr}/d$ and $\widetilde{l}_{r,corr}/d$ versus $Wi$ for both $\partial V_{\theta}/\partial r$ and $\partial V_{r}/\partial r$, respectively. }

\label{fig26}
\end{figure}

\subsection{Statistics of velocity gradients and their structure functions. \label{sub:Statistics}}

Further verification of the scaling laws in elastic turbulence in the channel flow comes from the statistical analysis of the velocity field. We point out that the statistical analysis of the longitudinal velocity component in spatial and in particular temporal domains show large scatter due to insufficient data that makes the analysis unreliable. On the other hand, the same analysis in a spatial domain of both components of the velocity gradients exhibits much better results.

First, we conducted the statistical analysis of the spatial increments of the radial gradients of the longitudinal velocity $\widetilde{\delta \partial V_{\theta}/\partial r}(\delta r/d)=[\partial V_{\theta}/\partial r (r/d+\delta r/d)-\partial V_{\theta}/\partial (r/d)]/\delta \partial V_{\theta}/\partial r (\delta r/d)_{rms}$ in a wide range of the spatial scales from 9.2 to 46 $\mu$m with the step 4.6 $\mu$m at $Wi=951$ for the bend $N=42$. The corresponding PDFs of the spatial increments of the normalized longitudinal velocity gradients $\widetilde{\delta {(\partial V_{\theta}/\partial r)}}(\delta r/d)$  have a small Gaussian cap and show well-pronounced exponential tails, clear scale invariance and symmetry with small scatter in spite of low statistics (see Fig. 27). Further analysis can be done in equivalent way either by direct calculations of the structure functions or by calculations of the moments of PDFs. The corresponding second moments of PDFs $S_{2,\theta}(\delta r/d)$ on $\delta r/d$ in log-log coordinates for several values of $Wi$ show the scaling region in $\delta r/d$ between 0.01 and 0.05 in Fig. 28. The structure functions of the higher even orders up to $p=8$ $S_{p,\theta}(\delta r/d)$  as a function of $\delta r/d$ are plotted in log-log coordinates at $Wi=951$ in Fig. 29. Due to symmetrical shape of PDFs odd moments are zero. The power dependence of the structure functions (or moments) $S_{p,\theta}(\delta r/d)\sim (\delta r/d)^{\zeta_{p,\theta}}$ is found in the range of $\delta r/d$ between 0.004 and 0.05. The plot in Fig. 30 demonstrates independence of the scaling exponents $\zeta_{2,\theta}$ and $\zeta_{4,\theta}$ of $Wi$ in the whole range of elastic turbulence. The dependence of $\zeta_{p,\theta}$  which are surprisingly close to the linear scaling with $\zeta_{p,\theta}=0.75p$ (see Fig. 31). The latter is very different from passive scalar behavior \cite{jun2}.
Analogous analalysis was conducted also for the radial gradients of the transversal velocity $\widetilde{\delta (\partial V_{r}/\partial r)}(\delta r/d)$ and the obtained results were very similar to the velocity gradient of the longitudinal component. The corresponding PDFs also exhibit similar features: small Gaussian cap, exponential tails though not so clean, scale invariance and symmetry (see Fig. 32). As the result, the second $S_{2,r}(\delta r/d)$ and higher order even $S_{p,r}(\delta r/d)$ moments (or structure functions) up to $p=8$  show scaling in the same range of scales (Figs. 33, 34, 35) with the scaling exponents $\zeta_{p,r}$ and $Wi$ (see Fig. 35) and $p$ (see Fig. 36) dependence close to those found for $\widetilde{\delta {(\partial V_{\theta}/\partial r)}}(\delta r/d)$.

\begin{figure}
\includegraphics[width=8cm]{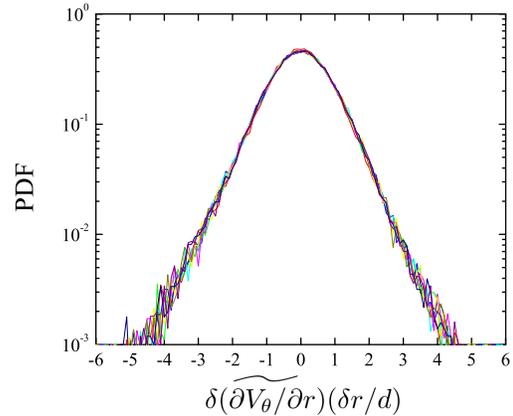}

\caption{PDFs  of the spatial increments of the normalized longitudinal velocity radial gradient $\widetilde{\delta (\partial V_{\theta}/\partial r}(\delta r/d))=[{\partial V_{\theta}/\partial r}(r/d+\delta r/d)-{\partial V_{\theta}/\partial r}(r/d)]/\delta({\partial V_{\theta}/\partial r}(\delta r/d))_{rms}$ at different length scales (from 9.2 to 46 $\mu$m with the step 4.6 $\mu$m) at $Wi=951$ based on its spatial velocity field measurements for the bend $N=42 $.}

\label{fig27}
\end{figure}

\begin{figure}
\includegraphics[width=8cm]{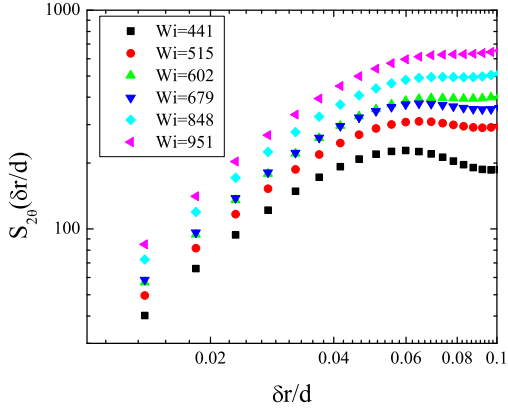}

\caption{Second moments $S_{2,\theta}(\delta r/d)$ of PDFs of the longitudinal velocity gradient increments $\widetilde{\delta (\partial V_{\theta}/\partial r}(\delta r/d))$ versus $\delta r/d$ for several values of $Wi$ (in log-log coordinates).}

\label{fig28}
\end{figure}

\begin{figure}
\includegraphics[width=8cm]{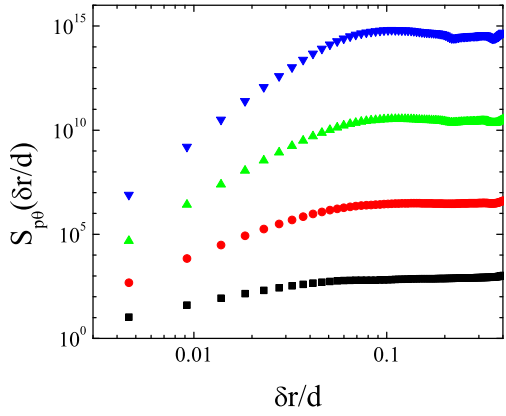}

\caption{Structure functions $S_{p,\theta}(\delta r/d)$ of the longitudinal velocity gradient spatial increments $\widetilde{\delta (\partial V_{\theta}/\partial r}(\delta r/d))$ up to $p=8$ (only even) for $Wi=951$ (in log-log coordinates).}

\label{fig29}
\end{figure}

\begin{figure}
\includegraphics[width=8cm]{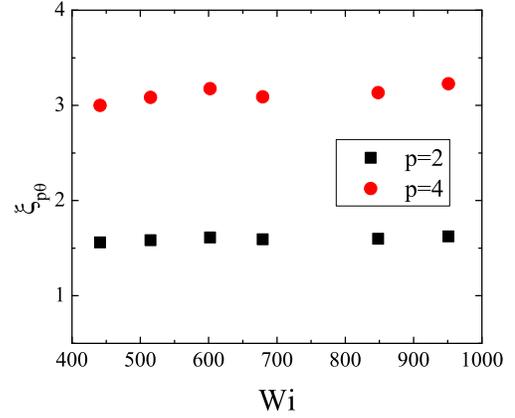}

\caption{Scaling exponents of the second $S_{2,\theta}(\delta r/d)$ and fourth $S_{4,\theta}(\delta r/d)$ moments of the longitudinal velocity gradient increments
$\widetilde{\delta (\partial V_{\theta}/\partial r}(\delta r/d))$ for various $Wi$.}

\label{fig30}
\end{figure}

\begin{figure}
\includegraphics[width=8cm]{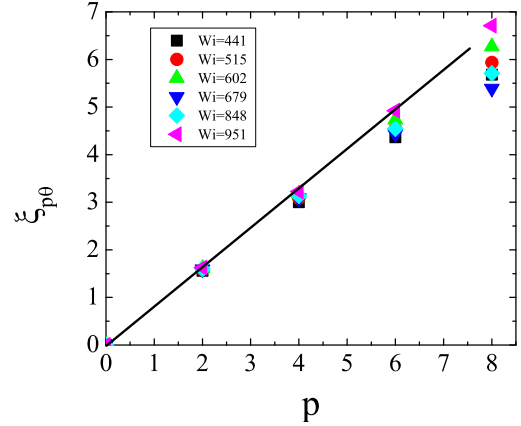}

\caption{ Scaling exponents $\zeta_{p,\theta}$ of the structure functions of the longitudinal velocity gradients spatial increments $\widetilde{\delta (\partial V_{\theta}/\partial r}(\delta r/d))$ versus $p$ for various $Wi$.}

\label{fig31}
\end{figure}

\begin{figure}
\includegraphics[width=8cm]{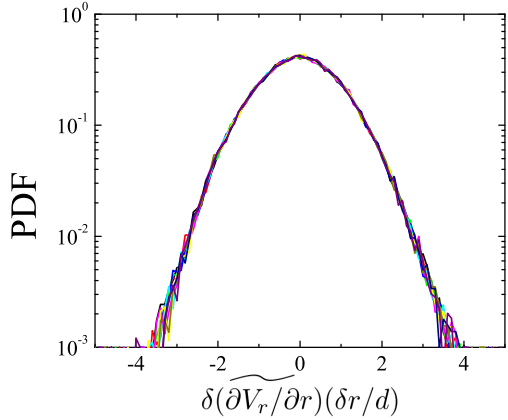}

\caption{PDFs  of the spatial increments of the normalized transversal velocity radial gradient $\widetilde{\delta (\partial V_{r}/\partial r}(\delta r/d))=[{\partial V_{r}/\partial r}(r/d+\delta r/d)-{\partial V_{r}/\partial r}(r/d)]/\delta({\partial V_{r}/\partial r}(\delta r/d))_{rms}$ at different length scales from 9.2 to 46 $\mu$m with the step 4.6 $\mu$m at $Wi=951$ based on its spatial velocity field measurements for the bend $N=42 $.}

\label{fig22}
\end{figure}

\begin{figure}
\includegraphics[width=8cm]{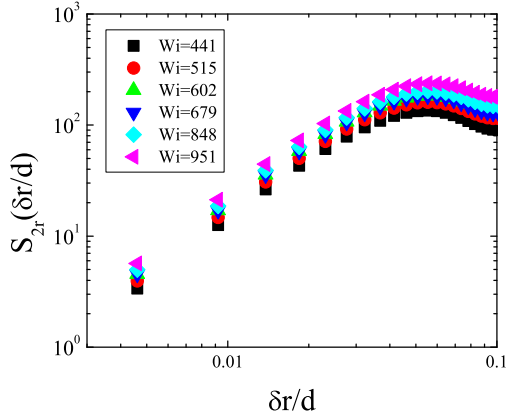}

\caption{Second moments $S_{2,r}(\delta r/d)$ of PDFs of the transversal velocity gradient increments $\widetilde{\delta (\partial V_{r}/\partial r}(\delta r/d))$ versus $\delta r/d$ for several values of $Wi$ (in log-log coordinates).}

\label{fig33}
\end{figure}

\begin{figure}
\includegraphics[width=8cm]{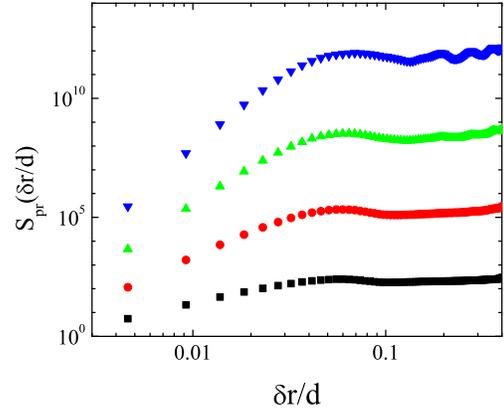}

\caption{Structure functions $S_{p,r}(r/d)$ of the transversal velocity gradient increments $\widetilde{\delta (\partial V_{r}/\partial r}(\delta r/d))$ up to $p=6$ (only even) for $Wi=951$ (in log-log coordinates).}

\label{fig34}
\end{figure}

\begin{figure}
\includegraphics[width=8cm]{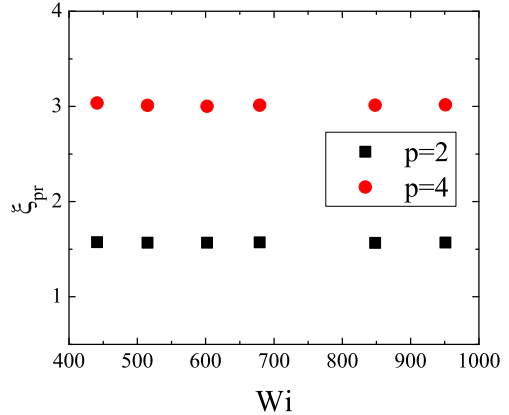}

\caption{Scaling exponents of the second $S_{2,r}(\delta r/d)$ and fourth $S_{4,r}(\delta r/d)$ moments of the transversal velocity gradient increments
$\widetilde{\delta (\partial V_{r}/\partial r}(\delta r/d))$ for various $Wi$.}

\label{fig35}
\end{figure}

\begin{figure}
\includegraphics[width=8cm]{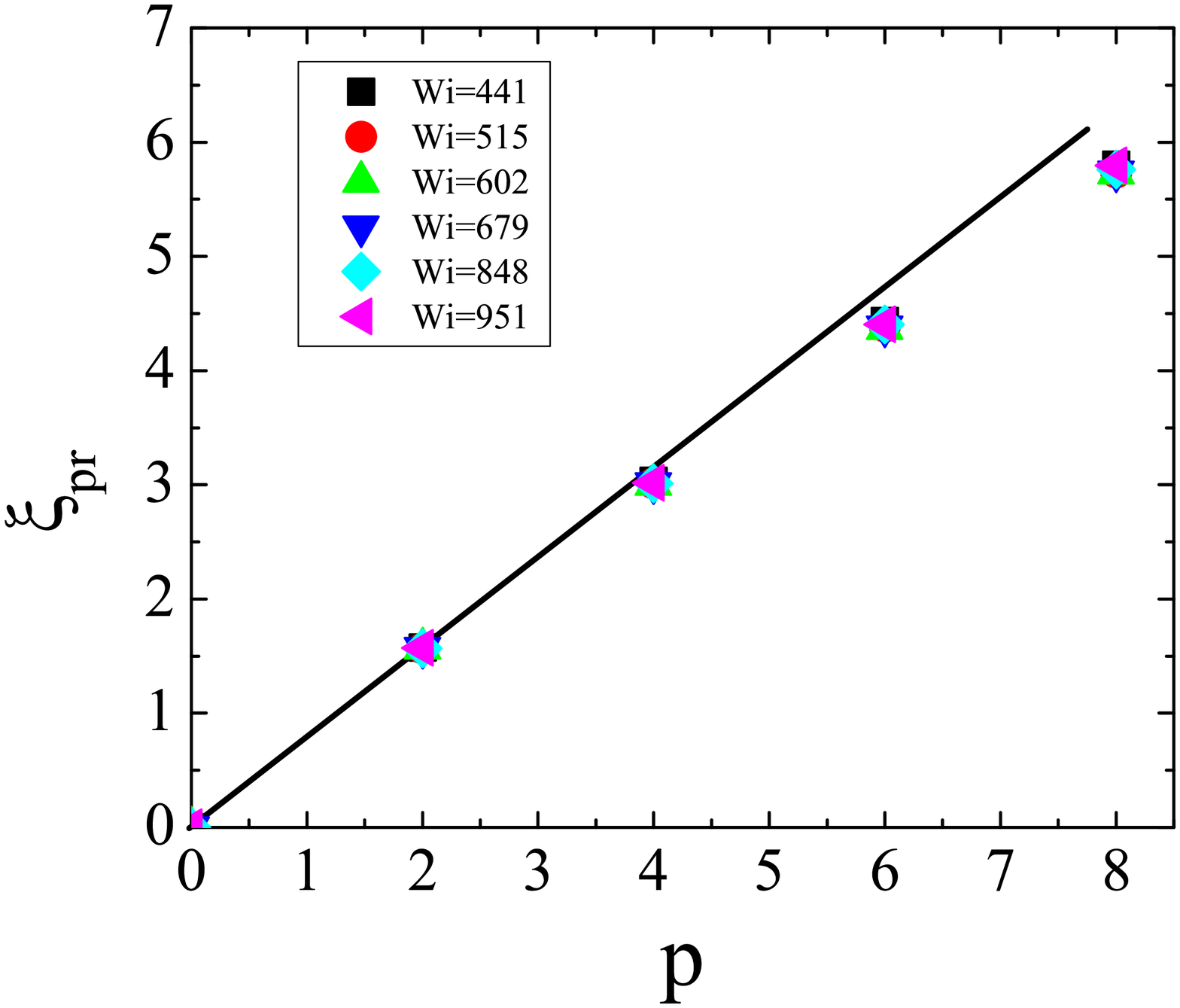}

\caption{ Scaling exponents $\zeta_{p,r}$ of the structure functions of the transversal velocity gradients increments $\widetilde{\delta (\partial V_{r}/\partial r}(\delta r/d))$ versus $p$ for various $Wi$.}

\label{fig36}
\end{figure}

\section{Discussion \label{sec:Discussion}}

Let us summarized first the main observations reported above.

(i) Well-defined threshold of the elastic instability in a curvilinear channel flow is identified by the average   longitudinal velocity and rms longitudinal and transversal velocity fluctuations (Figs. 4-6). The transition to elastic turbulence regime is determined from $V^{rms}_{\theta}$ and $V^{rms}_{r}$ as well as  $(\partial V_{\theta}/\partial r)^{rms}$, or $Wi_{loc}$, dependencies on $Wi$ (Figs. 5-7). In the latter case, an exponential  dependence of $Wi_{loc}$ on $Wi$ above the transition to elastic turbulence is observed (Fig. 7). (ii) The profiles of the average longitudinal velocity are altered drastically in elastic turbulence compared to a laminar flow (Fig. 9). In the elastic turbulence regime the characteristic boundary layer is clearly identified near the wall (Fig. 9 and 10). Being normalized and rescaled, all velocity profiles are collapsed on one curve with a horizontal flat part in the bulk and sharp change near the wall, so that all profiles have the same boundary layer width independent of $Wi$ (Figs. 18 and 19). (iii)  On the other hand, a profile of $(\partial V_{\theta}/\partial r)^{rms}$ (Fig. 14) shows  a peak much closer to the wall, inside the velocity boundary layer discussed above, which location near the channel wall is also independent of $Wi$. As we discussed in our early papers on elastic turbulence in a swirling flow \cite{teo2,teo3}, the peak in $(\partial V_{\theta}/\partial r)^{rms}$ means also the maximum in elastic stresses, and so the boundary layer is defined by the non-uniform spatial distribution of the elastic stresses across the channel. (iv) The correlation times determined from the temporal correlation functions for both velocity components differ up to 5 times. The correlation time for the transversal velocity component $\tau_{r,corr}$ is of the order of the polymer relaxation time, whereas the correlation time of the longitudinal velocity component $\tau_{\theta,corr}$ is several times larger (Figs. 21 and 22). (v) The correlation lengths found from the spatial correlation functions of both velocity components are of about $l_{corr}/d\approx 0.18$ for both velocity components in the elastic turbulence regime (Figs. 23 and 24). On the other hand, the correlation lengths obtained from the spatial correlation functions of the radial gradients of the longitudinal and transversal velocity components are about 6 times smaller than for the velocity components in the whole range of elastic turbulence (Figs. 25 and 26). (vi) PDFs of the spatial increments of the radial gradients of the longitudinal and transversal velocity components in a wide range of the length scales up to the boundary layer width demonstrate the scale invariance and exponential tails. It reminds very much the properties of the PDFs of passive scalar in the Batchelor regime \cite{jun2}. On the other hand, the second and higher order even moments of the PDFs, contrary to the passive scalar logarithmic dependence on spatial scales, show an algebraic increase with spatial increment with the scaling exponents $\tilde{\zeta}_{p,\theta}$ and $\tilde{\zeta}_{p,r}$, which dependence on $p$ mildly deviates from linear dependence.

There are several important messages, which follow from the observations summarized above.
(i) The elastic instability transition in a curvilinear channel is continuous one (forward bifurcation) as already found in early experiments \cite{NJP,teo1,teo4}, in contrast to those observed in Couette-Taylor flow \cite{NJP} and swirling flow between two disks \cite{NJP,teo3}. (ii) Contrary to the predictions, $Wi_{loc}$ grows with $Wi$ in the elastic turbulence regime and its value exceeds the theoretically predicted an order of unity by more than two orders of magnitude. (iii) The profiles of the average longitudinal component of the velocity reveal the boundary layer, which width is independent of $Wi$ in the whole range of elastic turbulence and being scaled by the channel width also independent on the characteristic size of the system. Both existence of the boundary layer and independence of its width on the control parameters of the flow are rather surprising features, in particular taking into account that the largest scale of the flow is smaller than the dissipation scale. This fact was already reported in Refs. \cite{teo2,teo3} for the swirling flow. (iv) The boundary layer observed in the average velocity profiles is a reflection of a nonuniform distribution of the rms of the radial gradient of the longitudinal velocity. Indeed, the latter profiles exhibit more intricate behavior with sharp peaks inside the velocity boundary layer, which locations are independent of $Wi$ but the peak values grow with $Wi$. Since $(\partial V_{\theta}/\partial r)^{rms}$ controls the degree of polymer stretching in a random flow and in this way the elastic stress, one concludes that similar nonuniform distribution of the elastic stress can be expected near the wall in elastic turbulence. It is a subject for future experiments. The expected nonuniform distribution of the elastic stress reminds a nonuniform distribution of passive scalar in a bounded channel flow \cite{teo4,jun2,chertkov1} with rare and strong ejection of jets occurring in the diffusion boundary layer and protruding into the peripheral region and even further into the bulk of the channel flow \cite{jun2,lebedev3,salman}. These jets are considerably alter mixing significantly reducing its efficiency. One can expect an emergence of similar jets of the elastic stresses, concentrated near the wall in the boundary layer, and ejected into the bulk. In such a way the elastic stresses are introduced into the flow. (v) Another characteristic spatial scale in the flow is the correlation length of the velocity field $l_{corr}$, which is about twice larger than the velocity boundary layer $l_{bl}$. Since elastic turbulence is a smooth random flow, where only a few large scale modes dominate the dynamics, one expects that $l_{corr}$ should be of the order of $d$. On the other hand, the correlation length defined from the correlation function of the velocity gradients $\tilde{l_{corr}}$ is about an order of magnitude smaller than $l_{corr}$. This fact points out on the relation of $\tilde{l_{corr}}$ with the characteristic spatial scale corresponding to the  peak location of $(\partial V_{\theta}/\partial r)^{rms}$ near the wall inside the velocity boundary layer width. Then due to eruption of jets of elastic stresses this characteristic scale is observed in the bulk. (vi) The same range of spatial scales are found in the scaling region of the structure functions of the velocity gradients $S_p$ that once more indicates a possible influence of jets protruding into the bulk of the flow.

\section{Conclusions \label{sec:Conclusions}}

To conclude, the experimental results show that one of the main predictions of the theory of elastic turbulence, namely the saturation of $Wi_{loc}$ in the bulk flow of elastic turbulence contradicts to the experimental observations both qualitatively and quantitatively in spite of the fact that the theory explains well the observed sharp decay of the velocity power spectrum \cite{lebedev,lebedev2}. The nonuniform distribution of $(\partial V_{\theta}/\partial r)^{rms}$ across the channel points out on the nonuniform distribution of elastic stresses. The latter may lead to the rare and strong eruption of the jets of elastic stresses from boundary layer into the bulk and in this way to introduce small spatial scales into the bulk flow. The existence of the resulting velocity boundary layer width and its peculiar properties require a proper theoretical description. Thus the experimental findings call for further development of theory of elastic turbulence in a bounded container, similar to what was done for a passive scalar problem \cite{chertkov1}.

\section*{Acknowledgments}
 This work is supported by grants from Israel Science Foundation and Lower Saxony Ministry of
Science and Culture Cooperation Grant.


\end{document}